%% file: berhault_PS.tex
\documentclass[10pt,twocolumn,twoside]{IEEEtran}
\usepackage{amsmath,graphicx}
\usepackage{amssymb}
\usepackage{setspace}
\usepackage{textcomp}
\usepackage{tikz,pgfplots}
\usetikzlibrary{plotmarks} 
\usepackage{color,colortbl}
\usepackage{stmaryrd}
\usepackage{wasysym}
\usepackage{paralist}

\newcommand{\im}{i }
\newcommand{\jm}{j }
\newcommand{\ig}{m }
\newcommand{\jg}{q }
\newcommand{\jd}{k }
\newcommand{\ts}{t }
\newcommand{\ipe}{x }
\newcommand{\jpe}{y }
\newcommand{\hu}{\hat{u} }
\newcommand{\hU}{\hat{U} }

   \graphicspath{{../fig/}{./}}

\ifCLASSINFOpdf
   \graphicspath{{../fig/}}
\else
\fi

\begin{document}
%
\title{Partial Sums Computation In Polar Codes Decoding}
%
%
%

\author{Guillaume~Berhault*, Camille~Leroux, Christophe~Jego, Dominique~Dallet\thanks{The authors are with the IMS Research Lab., University of Bordeaux, IPB ENSEIRB-MATMECA, 351 Cours de la Libération, 33405 Talence Cedex, France. (e-mail:firstname.lastname@ims-bordeaux.fr)}}

\maketitle
%
\input{berhault_PS_abstract}
%
%
%
%
%
%
%
%
\IEEEpeerreviewmaketitle
%
%
%
\input{berhault_PS_introduction}
\input{berhault_PS_sc_decoding}
%
\input{berhault_PS_partial_sum_computation}

%
\input{berhault_PS_conclusion_and_perspectives}
%
%
%
%
%
%
%
\appendices
%
\input{berhault_PS_appendix_tau}
%
%
%
%
%
\ifCLASSOPTIONcaptionsoff
  \newpage
\fi
%
%
%
%
%
\bibliographystyle{IEEEtran}
%
\bibliography{refs}

%
%

%

%
%
%




\end{document}

%% file: berhault_PS_abstract.tex

\begin{abstract}
Polar codes are the first error-correcting codes to provably achieve the channel capacity but with infinite codelengths. For finite codelengths the existing decoder architectures are limited in working frequency by the partial sums computation unit. We explain in this paper how the partial sums computation can be seen as a matrix multiplication. Then, an efficient hardware implementation of this product is investigated. It has reduced logic resources and interconnections. Formalized architectures, to compute partial sums and to generate the bits of the generator matrix $\kappa^{\otimes n}$, are presented. The proposed architecture allows removing the multiplexing resources used to assigned to each processing elements the required partial sums.
\end{abstract}

%% file: berhault_PS_introduction.tex

\section{Introduction}
\label{sec:introduction}
\begin{figure}[t]
\centering
\includegraphics[width=0.6\linewidth]{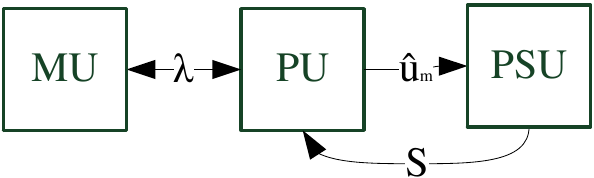}
\caption[Typical SC decoder structure]{Typical SC decoder structure.}
\label{fig:PU_MU_PSU}
\end{figure}

%
%
\IEEEPARstart{P}{olar} codes \cite{arikan_channel_2008} are a new class of error correction codes. These linear block codes are proven to achieve the capacity of any symmetric memoryless channel under successive cancellation (SC) decoding \cite{sasoglu_polarization_2009}. Nevertheless, they require a very large code length ($N=2^n>2^{20}$, \cite{arikan_channel_2008}) in order to actually approach the channel capacity. Consequently, the practical interest of polar codes highly depends on the possibility to design efficient encoder and decoder architectures for large codelengths.\\
%
%
When implemented in hardware (\cite{raymond_scalable_2013} and \cite{sarkis_fast_2013}), an SC decoder is composed of three main units: the \textit{processing unit} (PU), the \textit{memory unit} (MU) and the \textit{partial sums unit} (PSU) as seen in Fig. \ref{fig:PU_MU_PSU}. The decoded bits, $\hu_\ig$, are generated one after the other by the PU which needs
\begin{inparaenum}[(i)]
\item Log likelihood ratio (LLR) values ($\lambda$) stored in the MU, and
\item partial sums ($S$) calculated in the PSU
\end{inparaenum}. In SC decoding, the partial sums, which are used to carry on the decoding, are a combination of the previously decoded bits and are updated whenever a bit is decoded. \\
%
%
As shown in previous works \cite{leroux_semi-parallel_2012} and \cite{mishra_successive_2012}, the hardware implementation of SC decoders is constrained by the partial sums computation unit which occupies a major part of the area and limits the maximum working frequency, especially as $N$ grows. In \cite{zhang_low-latency_2013}, a method to compute partial sums is proposed but the best of our knowledge it has not been implemented. In \cite{berhault_partial_2013}, an efficient partial sum unit architecture was proposed and experimentally validated. However, no formal description of the concept has been given. The purpose of the paper is to bring some analytical contributions to \cite{berhault_partial_2013}. The proposed formalism could then be used with arbitrary kernels and extended to the structure of \cite{berhault_partial_2013}.\\

%% file: berhault_PS_sc_decoding.tex
\begin{figure}[h]
\centering
\includegraphics[width=0.7\linewidth]{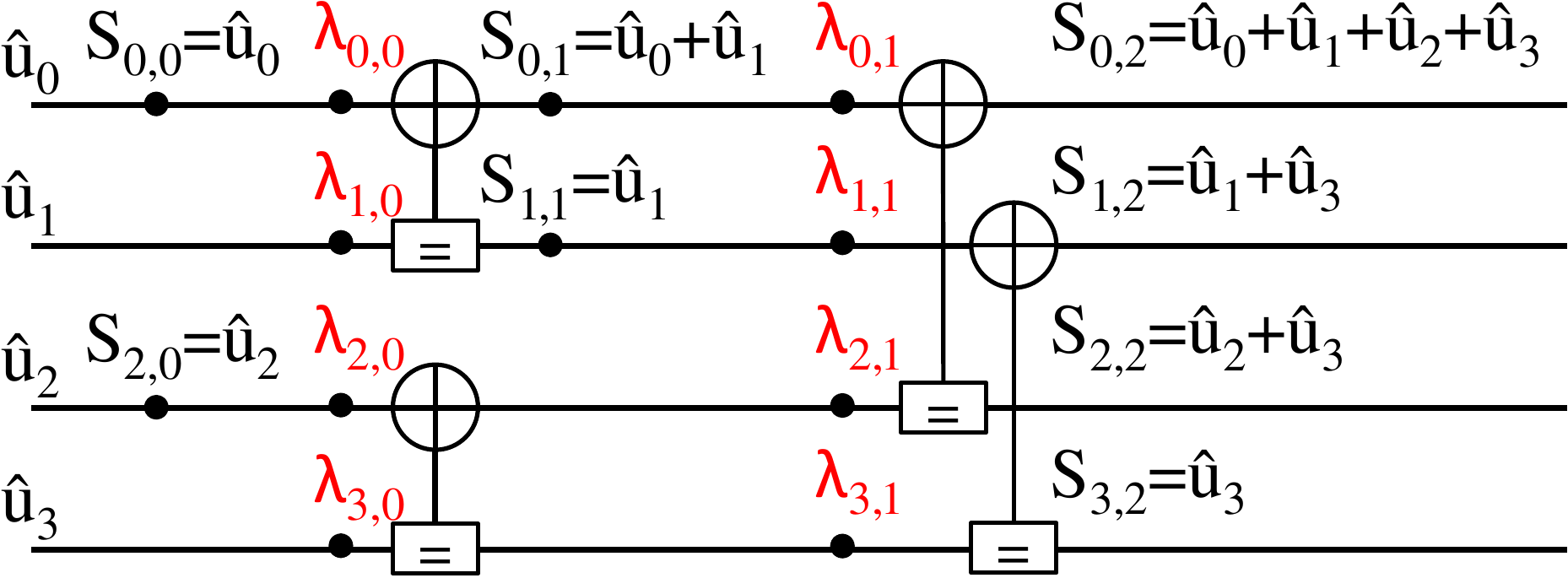}
\caption{Factor graph for $N=4$ polar code.}
\label{fig:factor_graph_N_4}
\end{figure}
\begin{figure}[b]
\begin{minipage}[b]{.48\linewidth}
  \centering
  \centerline{\includegraphics[width=0.7\linewidth]{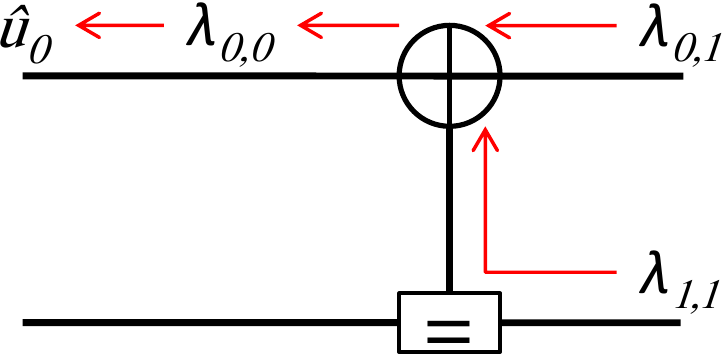}}
  \centerline{(a) Decoding of $\hat{u}_0$.}
\end{minipage}
\hfill
\begin{minipage}[b]{0.48\linewidth}
  \centering
  \centerline{\includegraphics[width=0.7\linewidth]{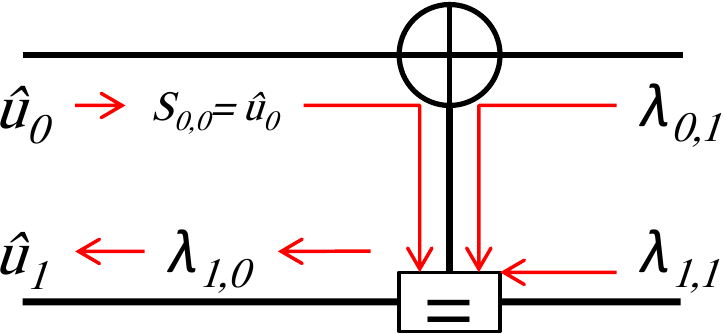}}
  \centerline{(b) Decoding of $\hat{u}_1$.}
\end{minipage}
\caption{$N=2$ polar code decoding example.}
\label{fig:Decoder_N_2}
\end{figure}
\section{Successive Cancellation Decoding}
\label{ssec:subpolarcodesdecoding}
For a code of length $(N=2^n)$, after being sent over the transmission channel, the noisy version $Y$ of the codeword $X$ is received. Each sample $y_\ig$ is converted into log likelihood ratio (LLR) format. These LLRs are denoted $\lambda_{\ig}$, with $0\leq \ig\leq N-1 $. The decoder successively estimates every bit $u_\ig$ based on the channel observation vector ($\lambda_0^{N-1}$) and the previously estimated bits ($\hat{u}_0^{\ig-1}$). In order to estimate each bit $u_\ig$, the decoder computes the following LLR value:
\begin{equation}
\lambda _{\ig,0} = \log\dfrac{\Pr(y_0^{N-1},\hat{u}_0^{\ig-1}|u_\ig=0)}{\Pr(y_0^{N-1},\hat{u}_0^{\ig-1}|u_\ig=1)}.
\label{eq:llr}
\end{equation}
The estimated bit $\hat{u}_\ig$ is calculated based on the following rule:
\begin{equation}
\hat{u}_\ig=\left\{\begin{array}{ll}0 & \text{if $\lambda _{\ig,0}>0$}\\
1 & \text{otherwise.}\end{array}\right.
\label{eq:u_\im_decision_rule}
\end{equation}
As proposed by Ar{\i}kan in \cite{arikan_channel_2008}, the factor graph representation of polar codes can be used to efficiently compute the $\lambda _{\ig,0}$. For a code of length $(N=2^n)$, the associated factor graph has $n$ columns and $N$ rows. SC decoding can be seen as an instance of belief propagation decoding where LLRs are propagated on the factor graph of the code with a particular scheduling. In SC decoding, bits $\hat{u}_\ig$ are processed sequentially and the decision is then fed back into the graph for the decoding of subsequent bits. In Fig. \ref{fig:Decoder_N_2}, the decoding on the factor graph of a simple $N=2$ polar code is represented. The graph is composed of a check node (CN or {$\oplus$}) and a variable node (VN or $\boxed{=}$). 
In general, the decoder successively estimates the bits $\hat{u}_\ig$ from the computation of LLRs of the indexed edges. The LLR of edge $(\ig,\jg)$ is computed such as:
\begin{equation}
\lambda _{\ig,\jg}= \left\{\begin{array}{l l}f\left (\lambda _{\ig,\jg+1},\lambda _{\ig+2^\jg,\jg+1}\right) &\text{if $B(\ig,\jg)=0$}\\
g\left(\lambda _{\ig-2^\jg,\jg+1},\lambda _{\ig,\jg+1},S_{\ig-2^\jg,\jg}\right) &\text{if $B(\ig,\jg)=1$,}\end{array}\right.
\end{equation}
with:
\begin{eqnarray}
\left\{\begin{array}{ll}
f(a,b) &= sgn(ab)\times \min(|a|,|b|)\\
g(a,b,s) &= b\oplus(-1)^s a.
\end{array}\right.
\label{eq:f_g_function}
\end{eqnarray}
where $B(\ig,\jg)\equiv \lfloor \dfrac{\ig}{2^\jg}\rfloor\text{ mod } 2$, $0\leq \ig < N\text{ and } 0\leq \jg < n$.
$S_{\ig,\jg}$ represents the \textit{partial sum}, located at the $\ig^\text{th}$ row and $\jg^\text{th}$ column of the factor graph. It corresponds to the propagation of decisions back into the factor graph. The partial sum set is denoted 
$$
\mathcal{S}=\{S_{\ig,\jg} | \ig\in \llbracket0;N-1 \rrbracket, \jg\in \llbracket0;n \rrbracket\}.
$$
The elements of the partial sum set $\mathcal{S}$ are not all used during the SC decoding, only those such as $B(\ig,\jg)=0$. For example $S_{2,1}=\hu_2+\hu_3$ is updated two times, when $\hu_2$ and $\hu_3$ are generated by the PU. 

%
%

%% file: berhault_PS_partial_sum_computation.tex
\section{From matrix product to register-based architecture}\label{sec:from-matrix-product-to-register-based-architecture}
\subsection{Matrix product representation}\label{sec:matrix-product-representation}
We now wish to prove that the set composed of the bits of $P_n(\ts)$, for $0\leq \ts < N$, contains all the elements of the set $\mathcal{S}$. Let us define the proposition $\mathcal{Q}_n$: \textit{\textquotedblleft All the partial sums of the factor graph are included in the set that contains the values of the vector $P_n(\ts)$, for $N=2^n$ and $0 \leq \ts < N$\textquotedblright, for all $n\in\mathbb{N}^*$.}\\
Let us verify that $\mathcal{Q}_1$ is true.\\
\begin{itemize}
\item when $\ts=0$\\
$P_1(0)=\hU_1(0)\times \kappa^{\otimes 1}=[\hu_0;0]\times \left[\begin{array}{cc}1&0\\1&1\end{array}\right]=[\hu_0;0]=[p_0(0) \, p_1(0)].$\\
One can notice that $p_0(0)=\hu_0=S_{0,0}$ as seen in Fig. \ref{fig:factor_graph_N_4}.\\
\item when $\ts=1$\\
$P_1(1)=\hU_1(1)\times \kappa^{\otimes 1}=[\hu_0;\hu_1]\times \left[\begin{array}{cc}1&0\\1&1\end{array}\right]=[\hu_0 \oplus \hu_1;\hu_1]=[p_0(1) \, p_1(1)].$\\
\end{itemize}
One can notice that $p_0(1)=\hu_0\oplus\hu_1=S_{1,0}=S_{1,1}$ and $ p_1(1)=\hu_1=S_{0,1}$ as seen in Fig. \ref{fig:factor_graph_N_4}.\\
The computations of $P_n(\ts)$ for $\ts=0$ and $\ts=1$ generate all the required partial sums to decode a code of size $n=1$. Therefore $\mathcal{Q}_1$ is true.\\
Assuming that, for $n\in \mathbb{N}^*$, $\mathcal{Q}_n$ is true, let us show that $\mathcal{Q}_{n+1}$ is true as well. Let us define two $N$-bit vectors:
\begin{itemize}
\item $\hat{V}_n(\ts)=\hu_0^{\ts}$ for $0 \leq \ts < N-1$, 
\item $\hat{W}_n(\ts)=\hu_N^{\ts}$ for $N \leq \ts < 2N-1$, 
\end{itemize} such as $\hU_{n+1}(\ts)=[\hat{V}_n(\ts) ; \hat{W}_n(\ts)]$.
 During the decoding of the $N$ first bits, $\hU_{n+1}(\ts)$ is equivalent to the concatenation of two $N$-bit vectors $\hat{V}_n(\ts)$ and $0_N$, such that $\hU_{n+1}(\ts)=[\hat{V}_n(\ts), 0_N]$. The matrix multiplication between $\hU_{n+1}(\ts)$ and $\kappa^{\otimes (n+1)}=\left[\begin{array}{cc}\kappa^{\otimes n}&0\\\kappa^{\otimes n}&\kappa^{\otimes n}\end{array}\right]$, for $\ts< N$, becomes:
\begin{equation}
P_{n+1}(\ts)=\hU_{n+1}(\ts) \times \kappa^{\otimes (n+1)}=[\hat{V}_n(\ts)\times \kappa^{\otimes n},0_N].
\label{eq:first_half}
\end{equation}
\begin{figure}[t]
\centering
\includegraphics[width=1\linewidth]{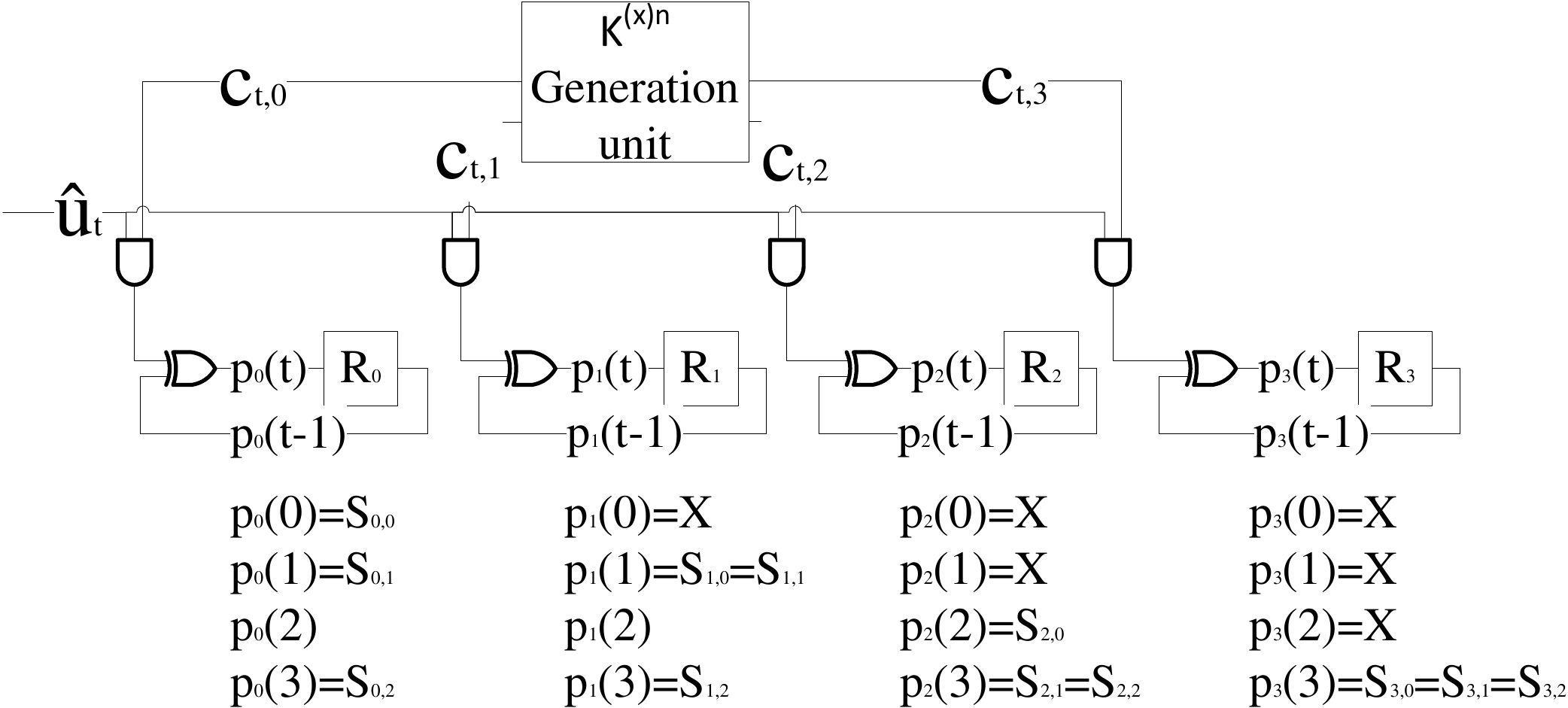}
\caption[Register-based architecture for partial sums computation]{Register-based architecture for partial sums computation ($N=4$).}
\label{fig:ps_element}
\end{figure}
Since $\mathcal{Q}_n$ is assumed true, all partial sums of the $N$ first rows and $n$ first columns of the factor graph are located in the $N$ leftmost bits of $P_{n+1}(\ts)$ ($0\leq \ts < N$) : $\hat{V}_n(\ts)\times \kappa^{\otimes n}$.\\
Similarly, during the decoding of the $N$ last bits, $\hU_{n+1}(\ts)$ is equivalent to the concatenation of two $N$-bit vectors $\hat{V}_n=\hat{V}_0^{N-1}$ and $\hat{W}_n(\ts)$, such that $\hU_{n+1}(\ts)=[\hat{V}_n, \hat{W}_n(\ts)]$. The matrix product between $\hU_{n+1}(\ts)$ and $\kappa^{\otimes (n+1)}$, for $ \ts\geq N $, becomes:
\begin{equation}
P_{n+1}(\ts)=[(\hat{V}_n\oplus\hat{W}_n(\ts))\times \kappa^{\otimes n},\hat{W}_n(\ts)\times \kappa^{\otimes n} ].
\label{eq:second_half}
\end{equation}
Since $\mathcal{Q}_n$ holds, every partial sum of the $N$ last rows and $n$ first columns of the factor graph is located in the $N$ rightmost bits of the resulting vector ($N\leq \ts < 2N$) : $\hat{W}_n(\ts)\times \kappa^{\otimes n}$. \\
Finally, when $\ts=2N-1$ the resulting vector of the product contains the partial sums of the last column of the factor graph. Therefore, $\mathcal{Q}_{n+1}$ is true. As a consequence, every partial sums of the factor graph of a code of size $n$ are generated by computing $P_n(\ts)$, for $0 \leq \ts < N$.

\subsection{Register-based structure}\label{sec:register-based-structure}
$P_n(\ts)$ is composed of $N$ bits $p_\jm(\ts)$, for $0 \leq \jm < N$. Each bit is the result of a matrix multiplication and can be rewritten as
$$
\begin{array}{lr}
p_\jm(\ts)=\sum _{l=0}^{\ts}\left(\hu_{l}\times c_{l,\jm}\right)\text{ (mod 2)} & \forall (\jm,\ts)\in \llbracket0;N-1 \rrbracket^2
\end{array}
$$
where $c_{\im,\jm}$ are the elements of the matrix $\kappa^{\otimes n}$.
This sum can be split into two finite sums. The first one for $l\in \llbracket 0;\ts-1\rrbracket$, and the second one for $l=\ts$, $l$ being the index of the sum of $p_\jm(\ts)$.
Therefore, the previous equation can be rewritten as:
\begin{equation}
p_\jm(\ts)=p_\jm(\ts-1) \oplus \hu_{\ts}\times c_{\ts,\jm}.
\label{eq:recurrentSeries}
\end{equation}
Equation (\ref{eq:recurrentSeries}) is a recurrent series which can be implemented by the register-based structure shown in Fig. \ref{fig:ps_element} for $N=4$. Since $P_n(\ts)$ is an $N$-bit vector, an $N$-bit register is required to store $p_\jm(\ts)$, for $0 \leq \ts < N$, along with $N$ XORs and $N$ ANDs elements.
Every $p_\jm(\ts)$, for $0 \leq \ts < N$, is stored in the $\jm^\text{th}$ DFF, $R_\jm$. One can notice that the partial sums of the $\jm^\text{th}$ row of the graph are computed by $p_\jm(\ts)$. Therefore, the partial sums, $S_{\ig,\jg}$, located on the $\ig^\text{th}$ row of the graph are successively stored in $R_\ig$.
\begin{figure}[t]
\centering
\includegraphics[width=0.8\linewidth]{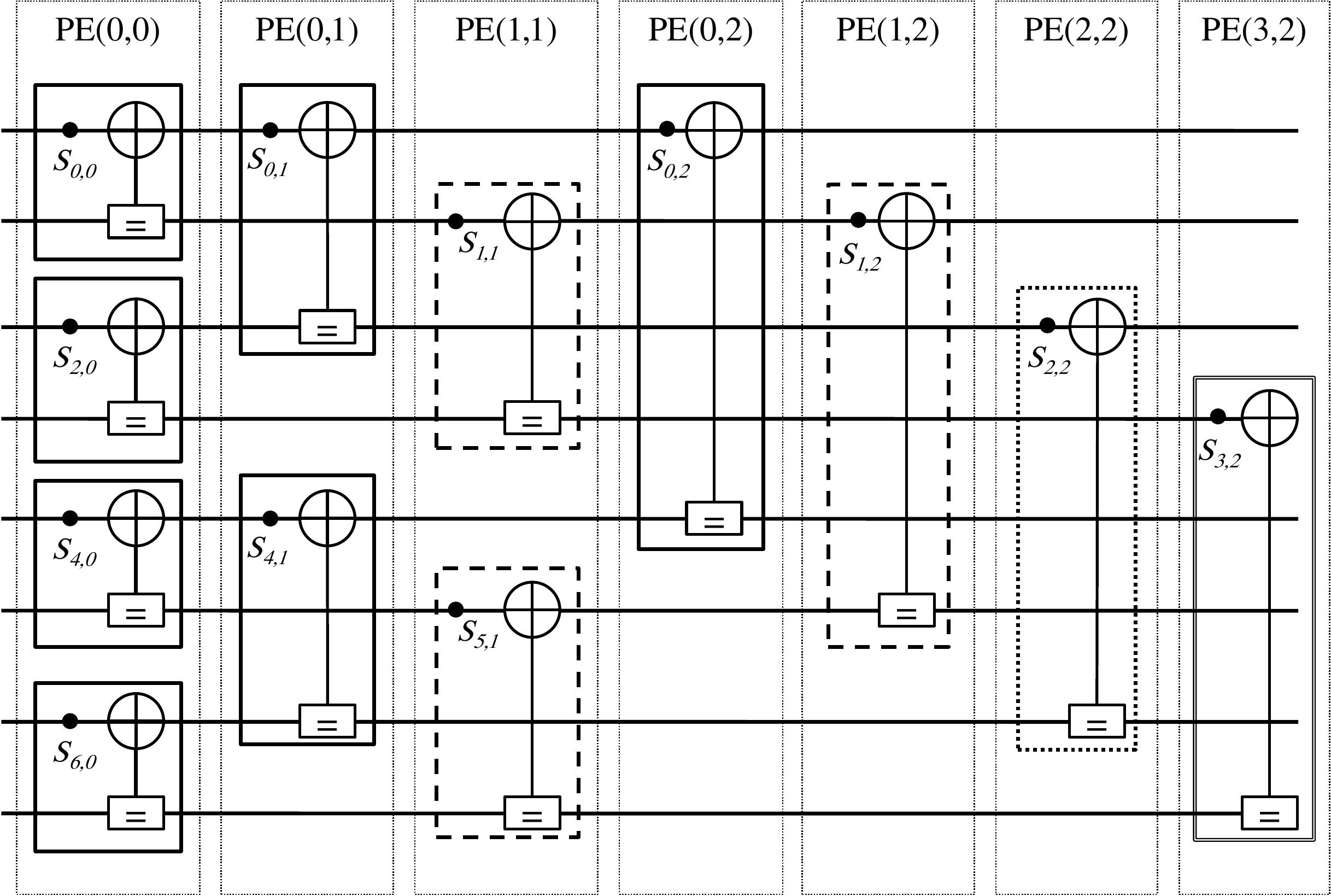}
\caption[Factor graph with identification to PE and the required partial sums ($N=8$)]{Factor graph with identification to PE and the required partial sums ($N=8$).}
\label{fig:PE_graph-crop}
\end{figure}
\section{Shift-register based structure}\label{sec:shift-register-based-partial-sums-generation}
In a tree SC decoder \cite{leroux_hardware_2011}, a PE can be assigned to the processing of one or more nodes in the graph. A PE is identified as PE($\ipe,\jpe$) such that $0 \leq \jpe \leq n-1$ and $0 \leq \ipe \leq 2^\jpe-1$. For instance, in Fig. \ref{fig:PE_graph-crop}, the partial sums $\{S_{0,0},S_{2,0},S_{4,0},S_{6,0}\}$ are assigned to PE$(0,0)$. Moreover, in the register-based architecture given in Fig. \ref{fig:ps_element}, the partial sum $S_{\ig,\jg}$ is stored in the DFF $R_\ig$. This means that a PE may be connected to multiple DFFs. Complex multiplexing resources are then necessary to select the partial sums for a given PE.\\
The main purpose of this section is to modify the PSU architecture detailed in Fig. \ref{fig:ps_element} so that all the partial sums required by a given PE are located in the same DFF. Such a structure would avoid any kind of multiplexing between a PE and the DFFs containing the required partial sums. 
\subsection{Partial sum location}\label{sec:partial-sum-location}
\begin{figure}[t]
\centering
\includegraphics[width=1\linewidth]{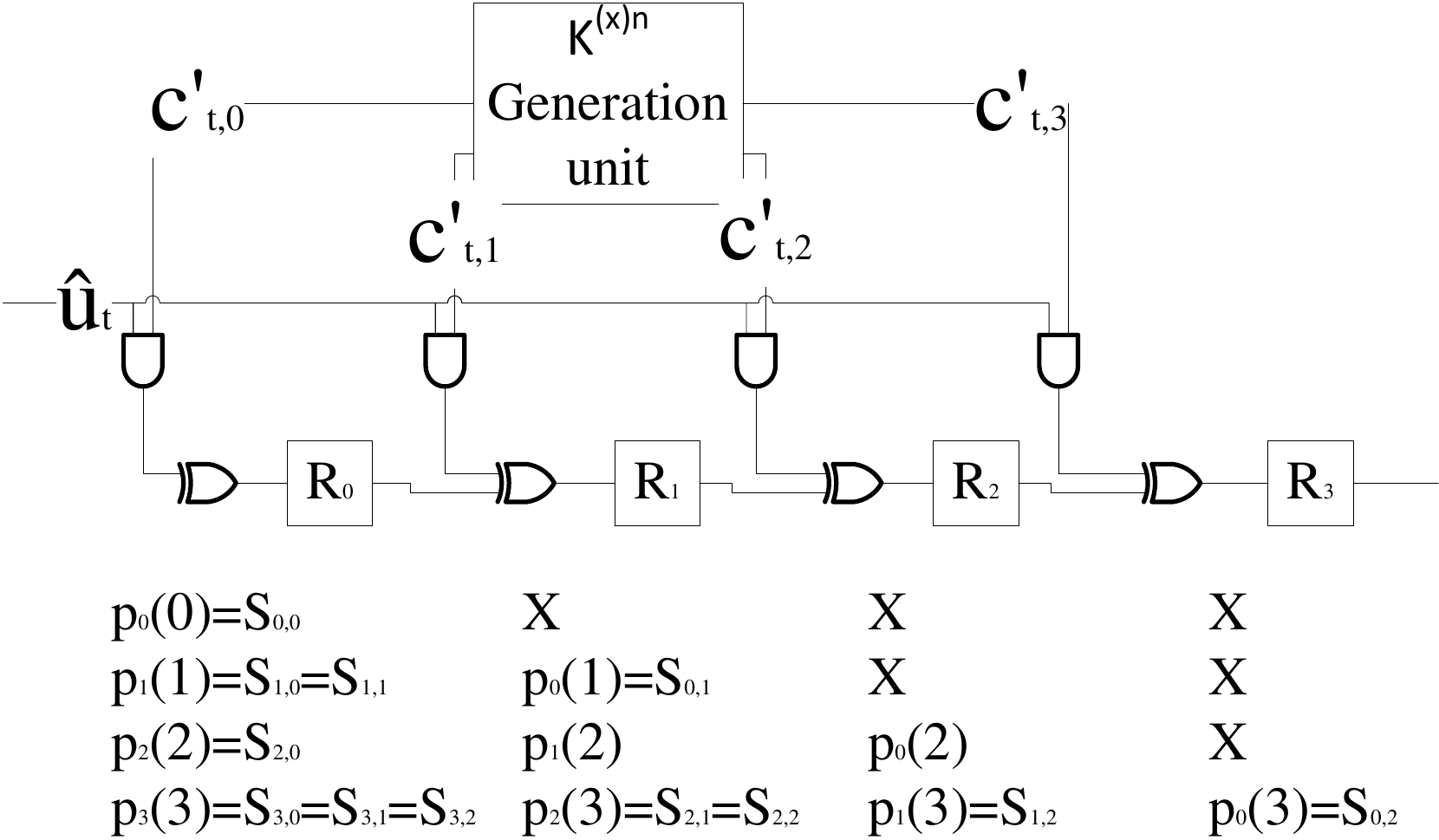}
\caption[Shifted structure of size N=4]{Shift-register-based architecture for partial sums computation (N=4).}
\label{fig:DFFShift-crop}
\end{figure}

The proposed structure is derived from the regular architecture depicted in Fig. \ref{fig:ps_element}. Instead of updating the current DFF value and store it back in the same DFF, it is possible to update and store this value in the next DFF as shown in Fig. \ref{fig:DFFShift-crop} for $N=4$.\\
The shift of the $p_\ig(\ts)$ values produces the exact same result as long as the coefficient of $\kappa^{\otimes n}$ are shifted accordingly. In this section we consider that the $c_{\im,\jm}$ bits are shifted as well in order to compute the same partial sums and are denoted $c'_{\im,\jm}$. Note that the generation of $\kappa^{\otimes n}$ is further detailed in section \ref{ssec:PScontrolGeneration}.\\
As shown in the previous section, without shift, the $\ig^\text{th}$ DFF contains the values $p_\ig(\ts)$, then the partial sum $S_{\ig,\jg}$. In the proposed architecture, due to the shift, $p_\ig(\ts)$ is not necessarily located in the $\ig^\text{th}$ DFF, thus neither is $S_{\ig,\jg}$. For example in Fig. \ref{fig:DFFShift-crop}, at time $\ts=0$, $p_0(0)$ is in $R_0$. At time $\ts=1$, $p_{1}(1)$ is in $R_{0}$ and $p_0(1)$ is in $R_1$. More generally, at time $\ts$, $p_\ig(\ts)$ is in $R_{\ts-\ig}$. This means that $S_{\ig,\jg}$ needs to be located, that is to say, one needs to determine the time of availability, $\tau$, such that $p_\ig(\tau)=S_{\ig,\jg}$. In APPENDIX \ref{sec:time-of-availability-for-a-partial-sum} it is shown that the partial sum $S_{\ig,\jg}$ is generated at time:
\begin{equation}
\tau=(\lfloor\frac{\ig}{2^\jg}\rfloor+1)\cdot2^\jg-1.
\label{eq:tau}
\end{equation}
In other words, at time $\tau$, the partial sum $S_{\ig,\jg}$ is located in the DFF $R_{\tau-\ig}$.

\subsection{DFF-PE direct connection}\label{sec:dff-pe-direct-connection}
It is now possible to know where and when any needed partial sum is located. However, the set of the partial sums that are required by a given PE has to be found in order to show that its elements are generated in the same DFF.\\
In Fig. \ref{fig:PE_graph-crop}, a PE($\ipe,\jpe$) requires all the partial sums that verify $S_{\ipe+k\cdot2^{\jpe+1},\jpe}$ with $k$ chosen such that $0 \leq \ipe+k\cdot2^{\jpe+1} \leq N-1$. For instance, PE($0,1$) requires $S_{0,1}$ ($k=0$) and $S_{4,1}$ ($k=1$). In the shift-register-based structure, the partial sum $S_{\ig,\jg}$ is located in the DFF $R_{\tau-\ig}$. This means that the set of partial sums required by PE($\ipe,\jpe$) are located in $R_{\tau-(\ipe+k2^{\jpe+1})}$. By replacing the expression of $\tau$, one can show that the set of DFF required by a PE($\ipe,\jpe$) are indexed by the expression $-(x \mod 2^\jpe)+2^\jpe-1$.
This index is independent of $\jd$. In other words, the partial sums required by PE($\ipe,\jpe$) are all located in the same DFF.
\\
Moreover, as $0\leq\jpe\leq n-1$, therefore $0\leq2^\jpe\leq \frac{N}{2}$. With these considerations, the previous expression of the DFF index ranges from $0$ to $\frac{N}{2}-1$ ($0 \leq - (x \mod 2^\jpe)+2^\jpe-1 \leq \frac{N}{2}-1$).
As a consequence, the $\frac{N}{2}$ first DFFs are sufficient to memorize all the required partial sums during the decoding of code of length $N$.\\
The proposed architecture can easily be applied to line SC decoders by grouping the PE which are assigned to the same DFFs (\cite{leroux_hardware_2011}). The shift-register-based architecture may also be employed for a semi-parallel SC decoder architecture by adding multiplexing.

\section{$\kappa^{\otimes n}$ matrix generation unit}
\label{ssec:PScontrolGeneration}
\begin{figure}[t]
\centering
\includegraphics[width=1\linewidth]{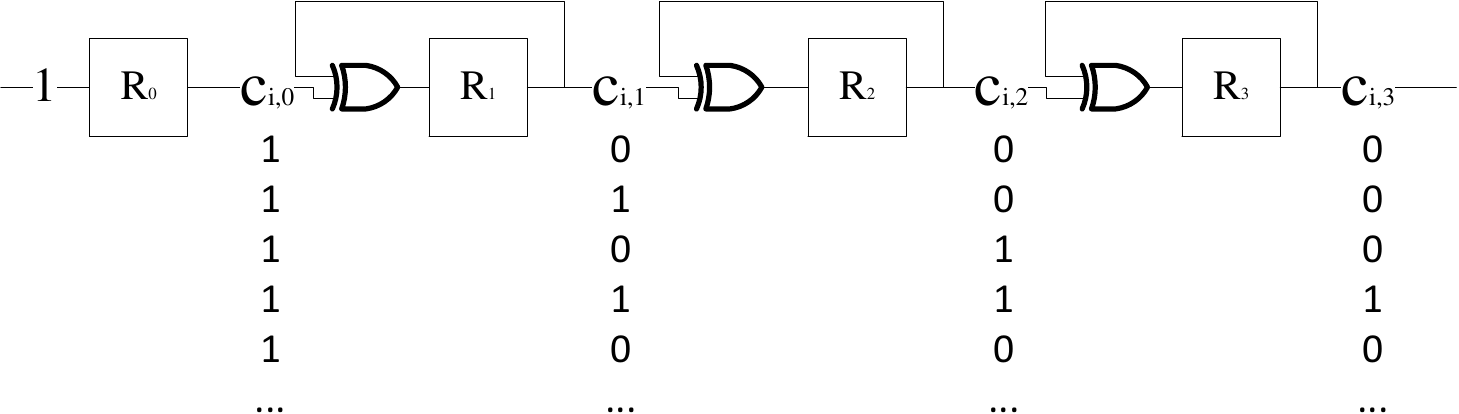}
\caption[]{$\kappa^{\otimes 2}$ matrix generation for a code of size $N=8$.}
\label{fig:control_bit_gen_N_8-crop}
\end{figure}

The partial sums calculations, for a code of length $N=2^{n}$, require the values of $\kappa^{\otimes (n-1)}$ two times as seen in equations (\ref{eq:first_half}) and (\ref{eq:second_half}), in section \ref{sec:from-matrix-product-to-register-based-architecture}, but for a code size of ($2^{n+1}$) instead of $2^n$. The first time to calculate the partial sums of the first half of the rows in the graph. The last one is for the remaining partial sums. The generation of the bits of the rows of $\kappa^{\otimes n}$ can be seen as a finite state machine with as many state as there are rows to generate. Each state represents a row of the matrix. Every row could be stored in a ROM but this architectural solution would become impractical for code length reaching $2^{20}$ bits. Another approach is to compute the value of the future state using the current state value. To apply this proposition, a quick observation of the matrix $\kappa^{\otimes (n-1)}$ is necessary. Two main properties can be highlighted:
%
$$
\begin{array}{cclll}
c_{\im,0} &=& 1 & \forall \ \im &\in \llbracket 0;\frac{N}{2}-1\rrbracket \\\\
c_{\im,\jm} &=& c_{\im-1,\jm-1} \oplus c_{\im-1,\jm} & \forall (\im,\jm) &\in \llbracket 1;\frac{N}{2}-1\rrbracket ^2 \\
\label{eq:prop}
\end{array}
$$
The first property means that the first bit is always one, which is immediate due to the Kronecker power definition. Therefore, this bit does not require recalculations when changing state.
The second property is the most important because it is exploited to compute the future state from the current state bit values.
\\
The rows of $\kappa^{\otimes (n-1)}$ are generated one after the other. Therefore, in $c_{\im,\jm}$, the index $\im$ represents the time $\ts$, while $\jm$ corresponds to the DFF index, in which the value is stored. The second property is the equation of the construction and can be rewritten as $M_\jm(\ts)=M_{\jm-1}(\ts-1) \oplus M_\jm(\ts-1)$. To implement such an equation, an AND logic gate and a DFF are sufficient. The $\frac{N}{2}$ DFFs are connected one to the other as seen in Fig. \ref{fig:control_bit_gen_N_8-crop}.\\
The shift-register based structure, used to compute the partial sums, requires that the bits of $\kappa^{\otimes n}$ are shifted accordingly. One can verify that the diagonals and the columns are equal. Therefore, the proposed structure generates the bits of $\kappa^{\otimes n}$ that can be employed as they are for the shift-register based architecture.

%% file: berhault_PS_conclusion_and_perspectives.tex

\section{Conclusion and perspectives}
\label{sec:conclu}
Designing efficient hardware decoders for polar codes would result in their potential inclusion in future digital telecommunication standards. State of the art works propose efficient successive cancellation decoder hardware designs whose limiting element is the partial sum unit. This paper brings contribution to formalizing the structure proposed in \cite{berhault_partial_2013}, reducing the hardware complexity. The shift-register based architecture can be extended to line SC decoder. It can also be applied to semi-parallel architectures by adding more multiplexing resources.\\
The proposed computation method opens the way for additional works such as the extension of this architecture to higher kernels or the enhancement of parallelism in this structure.

%% file: berhault_PS_appendix_tau.tex
\section{Time of availability for a partial sum}\label{sec:time-of-availability-for-a-partial-sum}
Any partial sum, $S_{\ig,\jg}$, can be seen as an element of a sub-codeword $SCW(\ig,\jg)$. This sub-codeword is the encoded version of the sub-code $\hu_a^b\in \hU$. All the elements of $SCW(\ig,\jg)$ are valid whenever all bits of $\hu_a^b$ are valid. Since the bits are decoded sequentially, a partial sum $S_{\ig,\jg}$ is valid when the bit $\hu_b$ is available, that is to say when $\tau=b$. The main purpose is then to find the expression of $b$. We already know $\jg$, thus $a$ is the only remaining variable to find before getting the expression of $b$. The following equality comes from the length of $SCW(\ig,\jg)$, which is the same as the length $\hu_a^b$.
\begin{equation}	
2^\jg = b-a+1.
\label{eq:subcode length}
\end{equation}
Since $a$ is the starting index, it is a multiple of $2^\jg$. The following expression returns the value of $a$:
\begin{equation}	
a=\lfloor\dfrac{\ig}{2^\jg}\rfloor*2^\jg.
\label{eq:starting index}
\end{equation}	
Now the only remaining variable is $b$. Using equation (\ref{eq:subcode length}) and (\ref{eq:starting index}) it follows:
$$
b=2^\jg-1+\lfloor\dfrac{\ig}{2^\jg}\rfloor*2^\jg.
$$
Finally, $S_{\ig,\jg}$ is only valid when $\tau=b$, that is to say:
\begin{equation*}
\tau=b=(\lfloor\dfrac{\ig}{2^\jg}\rfloor+1)*2^\jg-1.
\label{eq:tau expression}
\end{equation*}